\documentclass{article}
\usepackage[T1]{fontenc}
\usepackage{wrapfig}
\usepackage[portuges,english]{babel}
\usepackage{amssymb,latexsym,amsmath,color,mathrsfs,ifsym,graphics,stmaryrd}
\usepackage{amsthm, amsfonts, mathrsfs}
\usepackage[colorlinks,linkcolor=blue,urlcolor=blue,citecolor=black,
plainpages=false,pdfpagelabels,breaklinks]{hyperref}
\usepackage[all]{xy}
\usepackage{graphicx,caption}
\usepackage[margin=2.3 cm]{geometry}

\newcommand{\Tr}[0]{\operatorname{Tr}}
\newcommand{\bra}{\rangle}
\newcommand{\ket}{\langle}

\newcommand{\fig}{\linebreak\refstepcounter{figure}{\small Fig. \thefigure}}

\newtheorem{theorem}{Theorem}[section]
\newtheorem{definition}[theorem]{Definition}

\begin{document}

\title{{\sc The Many Inconsistencies of the\\Purity-Mixture Distinction in\\Standard Quantum Mechanics}}

\author{{\sc C. de Ronde}$^{1,2,3}$ and {\sc C. Massri}$^{4,5}$}
\date{}

\bibliographystyle{plain}
\maketitle

\begin{center}
\begin{small}
1. Philosophy Institute Dr. A. Korn, University of Buenos Aires - CONICET\\
2. Center Leo Apostel for Interdisciplinary Studies\\Foundations of the Exact Sciences - Vrije Universiteit Brussel\\
3. Institute of Engineering - National University Arturo Jauretche\\
4. Institute of Mathematical Investigations Luis A. Santal\'o, UBA - CONICET\\
5. University CAECE
\end{small}
\end{center}

\bigskip

\begin{abstract}
\noindent The distinction between {\it pure states} and {\it mixed states} is a kernel ingredient of what is considered to be the standard formulation of quantum mechanics and plays today a kernel role in foundational debates about the meaning of quantum probability, the separability of quantum systems, the definition and measure of entanglement, etc. In this work we attempt to expose the many inconsistencies introduced by this distinction and the serious consequences this has for many ongoing research programs within quantum physics which apply these notions uncritically.    
\end{abstract}
\begin{small}

\end{small}

\bigskip

\bigskip

\bigskip

\bigskip

\section{Quantum Theory: From Matrices to Vectorial Spaces}

The origin of Quantum Mechanics (QM) goes back to the year 1900 when Max Planck introduced the {\it quantum of action}, and consequently, the non-classical idea of {\it discrete} energy. After the successful applications of the quantum postulate to the explanation of the photoelectric effect by Albert Einstein in 1905 and the development of the Hydrogen model of the atom by Niels Bohr in 1913, Werner Heisenberg was able to develop a closed, consistent and coherent mathematical formalism capable to account in an invariant manner for the line intensities that were actually observed in the lab. Born and Jordan recognized that the tables of intensive data used by Heisenberg were in fact {\it matrices} and the strange multiplication rule an exposure of their non-commutative structure. Indeed, in his search for describing quanta, Heisenberg had re-discovered a well known field of mathematics, namely, matrix algebra. So let us begin by introducing some concepts and definitions about matrices.
A $n\times n$ complex \emph{matrix} is an array of $n$ by $n$ complex numbers. An example of a $2\times 2$ real matrix is
\[
\begin{pmatrix}
1&3\\2&-1
\end{pmatrix}
\]
It is usual to denote the space of complex $n\times n$
matrices as $\mathbb{C}^{n\times n}$ and the space of real $n\times n$ matrices as $\mathbb{R}^{n\times n}$. The entry in row $i$ and column $j$ of a matrix $[A]\in \mathbb{C}^{n\times n}$ is denoted as $[A]_{ij}=a_{ij}$. An interesting property of matrices is that 
the operations on complex (or real) numbers can be extended. For example, if $[A],[B]\in \mathbb{C}^{n\times n}$ are two matrices, where $[A]_{ij}=a_{ij}$ and $[B]_{ij}=b_{ij}$, we define the addition $[A]+[B]$ and multiplication $[A][B]$ as,
\[
([A]+[B])_{ij}:=a_{ij}+b_{ij},\quad
([A][B])_{ij}:=\sum_{k=1}^n a_{ik}b_{kj}. 
\]
As mentioned before, an essential aspect of matrices is that $[A][B]\neq [B][A]$. For example,
\[
\begin{pmatrix}
0&1\\
0&0
\end{pmatrix}
\begin{pmatrix}
1&0\\
0&0
\end{pmatrix}
=
\begin{pmatrix}
0&0\\
0&0
\end{pmatrix}
\quad
\neq
\quad
\begin{pmatrix}
0&1\\
0&0
\end{pmatrix}
=
\begin{pmatrix}
1&0\\
0&0
\end{pmatrix}
\begin{pmatrix}
0&1\\
0&0
\end{pmatrix}.
\]
One of that main applications of matrices in $\mathbb{C}^{n\times n}$  
is to encode linear transformations between $\mathbb{C}^n$ and $\mathbb{C}^n$.
Specifically, 
given a matrix $[A]$, we can define the linear transformation $f$ that sends the canonical $i$-vector, $|e_i\rangle=(0,\dots,0,1,0,\dots,0)$, to the vector 
$|c_i\rangle=(a_{1i},\dots,a_{ni})$ corresponding to the $i$-column of $[A]$. In this way, matrix multiplication corresponds to  composition of linear transformations. 

 
Heisenberg had developed QM in July 1925 by giving up the classical space-time representation inherited from modern classical physics and following instead the guide presented by Ernst Mach's observability principle ---which had also helped Einstein to develop his special theory of relativity. It is through this radical shift that he was able to focus in the intensive patterns that were already observed in the lab by experimentalists. As a result, the new mathematics proposed by Heisenberg ---completed jointly with Max Born and Pascual Jordan--- abandoned the infinitesimal calculus that physicists had learned to work with for more than two centuries creating an obvious tension with the {\it continuous} modern representation of physics in terms of `particles' and `waves' evolving within space and time. Of course, all physicists were happy to have reached a closed and consistent mathematical formalism that united the quantum experience observed in the lab (Appendix A) but ---regardless of the growing influence of positivism within physics--- the lack of an {\it anschaulicht} spatiotemporal understanding of quantum phenomena made them also quite uncomfortable. However, before they could get accustomed to the new mathematical framework, only six months later in January of 1926, Erwin Schr\"odinger sent a paper to {\it Annalen der Physik} where, taking de Broglie's ideas as a standpoint,\footnote{In 1924, the young French physicist Luis de Broglie had proposed a dualistic theory for describing atomic phenomena in terms of waves and particles. His idea was to generalize Einstein's theory of light quanta so that the motion of any particle had a wave associated with it as well. Where Einstein had assigned the properties of particles (light quanta) to the light wave (radiation) back in 1905, de Broglie now assigned the properties of radiation to particles. Einstein strongly supported de Broglie's ideas, but Pauli and Heisenberg did not. The attack of the two young friends was mercifulness and very soon even de Broglie would became himself dissatisfied with his theory.} he proposed a non-relativistic wave like differential equation that would give the correct energies for the quantized Hydrogen atom. In his attempt to combine a wave formulation with de Broglie's ideas, Schr\"odinger had arrived to the following differential equation:
\[
i \hbar\frac{d}{dt}\psi(t) = \hat{H} \psi(t)
\]
For obvious reasons, $\psi(t)$ became to be known as the ``quantum wave function''. Physicists were instantly captivated by the new formulation in terms of a differential equation which promised to restore the intuitive content to the theory through the provision of a {\it continuous} mathematical representation that would pave the road back to a spatiotemporal description of quantum phenomena. However, very soon it would become evident that the enthusiasm of the physics community was completely unjustified. Schr\"odinger himself would realize that the quantum wave equation did not represent a physical wave in 3-dimensional space-time. Instead, the domain of $\psi(t)$ was the multidimensional {\it configuration space} already known to physicists in the Hamilton-Jacobi formulation of classical mechanics. As described by Schr\"odinger:
\begin{quotation}\noindent {\small
``Above all, however, the departure from classical mechanics in the two theories seem to occur in diametrically opposed directions. In Heisenberg's work the classical continuous variables are replaced by systems of discrete numerical quantities (matrices), which depend on a pair of integral indices, and are defined by \emph{algebraic} equations. The authors themselves describe the theory as a `true theory of a {\it discontinuum}'. On the other hand, wave mechanics shows just the reverse tendency; it is a step from classical point-mechanics towards a \emph{continuum-theory}. In place of a process described in terms of a finite number of dependent variables occurring in a finite number of total differential equations, we have a continuous \emph{field-like} process in configuration space, which is governed by a single \emph{partial} differential equation derived from a principle of action.'' \cite{Sch5}}
\end{quotation}

A few months later after Schr\"odinger's wave formulation was presented, Max Born would provide a probabilistic interpretation of the quantum wave function where the introduction of a new discrete ``quantum jump'' ---similar to the one Bohr had proposed in his model---  would not only destroy the causal evolution of $\psi (x)$ but also reintroduce the reference to corpuscles. In his paper titled {\it On the Quantum Mechanics of Collision Processes}, Born interpreted $\psi (x)$ as encoding a probability density function for an elementary particle to be found at a given region. The wave function, a complex-valued function, had physical meaning only in terms of square modulus, $|\psi(x)|^2 = \psi (x)^* \psi (x)$. If $|\psi (x)|^2$ has a finite integral over the whole of three-dimensional space, then it is possible to choose a normalizing constant and the probability that a particle is within a particular region V is the integral over V of $|\psi (x)|^2$. In this way, Born was creating ---like Bohr had done before--- an irrepresentable bridge between a weird probability-wave and single `clicks' observed in detectors interpreted, in turn, as the ``self-evident'' effect of quantum particles. In this way, Born's rule was imposing a radical shift in the reference of quantum probability from intensive values ---which Heisenberg had originally considered as the observational standpoint in order to develop QM--- to single measurement outcomes. As remarked by Born \cite{Born26}: ``Schr\"{o}dinger's quantum mechanics [therefore] gives quite a definite answer to the question of the effect of the collision; but there is no question of any causal description. One gets no answer to the question, `what is the state after the collision' but only to the question, `how probable is a specified outcome of the collision'.'' Of course, even though the reference to particles and collisions was explicit within Born's narrative there was ---like in the case of Bohr's model of the atom--- no theoretical representation of them.

Already in 1926, Schr\"odinger himself would reach the conclusion that the matrix and wave formulations were mathematically equivalent. But in fact, the Austrian physicist had only demonstrated that his wave mechanics implied the matrix formulation, the converse was still missing. Later on, Wolfgang Pauli, Carl Eckart and John von Neumann would give  their own ---also incomplete--- proofs giving rise to what Fred Muller has termed ``the equivalence myth'' \cite{Muller97}. So even though all proofs were full of dead ends, the equivalence  between the two formalisms would become an accepted gospel between physicists. The idea that both mathematical formalisms where completely equivalent was given even more force in the year 1930 when the young English engineer and mathematician Paul Maurice Dirac would provide, through his transformation theory, a new axiomatic formulation of the theory of quanta in purely vectorial terms. In his book, {\it The Principles of Quantum Mechanics}, leaving Heisenberg's matrices behind, Dirac would link vectors with {\it quantum states} \cite{Dirac74}. Let us recall that a unit vector in Dirac notation can be written as a ket $|x\rangle$. Of course, this shift, from matrices to vectors, was already implicit in Schr\"odinger's equation, which could be naturally read as a vectorial equation. Following Dirac's notation, this could be seen as the time evolution of a vector $|v_t\rangle$ in a multi-dimensional vectorial space. 
\[
i \hbar\frac{d}{dt}|v_t\rangle = \hat{H}|v_t\rangle.
\]
Two years later, in 1932, John von Neumann would provide a more rigorous mathematical framework for Dirac's axioms through the introduction of Hilbert spaces. As presented by von Neumann himself in his famous book: 
\begin{quotation}
\noindent {\small ``The object of this book is to present the new quantum mechanics in a unified representation which, so far as it is possible and useful, is mathematically rigorous. This new quantum mechanics has in recent years achieved in its essential parts what is presumably a definitive form: the so-called `transformation theory.' Therefore the principal emphasis shall be placed on the general and fundamental questions which have arisen in connection with this theory. In particular, the difficult problems of interpretation, many of which are even now not fully resolved, will be investigated in detail. In this connection, the relation of quantum mechanics to statistics and to the classical statistical mechanics is of special importance.'' \cite[p. 1]{VN}}
\end{quotation}  

One can provide a definition of Hilbert spaces and dimension using sets and functions. A \emph{complex vector space} ${\cal H}$ can be regarded as a set with two algebraic structures satisfying the following axioms:
\begin{enumerate}
\item $x+(y+z)=(x+y)+z,\quad\forall x,y,z\in\mathcal{H}$
\item $x+y=y+x,\quad\forall x,y\in\mathcal{H}$
\item There exists $0\in\mathcal{H}$ such that $0+x=x,\quad\forall x\in\mathcal{H}$
\item There exists $-x\in\mathcal{H}$ such that $x+(-x)=0,\quad\forall x\in\mathcal{H}$.
\item $\lambda(x+y)=(\lambda x)+(\lambda y),\quad\forall \lambda\in\mathbb{C}, x,y\in  {\cal H}$.
\item $(\lambda_1+\lambda_2) x= (\lambda_1x)+(\lambda_2 x),\quad \forall \lambda_1,\lambda_2\in\mathbb{C}, x\in  {\cal H}$.
\item $(\lambda_1\lambda_2) x= \lambda_1(\lambda_2 x),\quad \forall \lambda_1,\lambda_2\in\mathbb{C}, x\in {\cal H}$.
\item $1x=x,\quad\forall x\in {\cal H}$.
\end{enumerate}
The most common example of a complex vector space is $\mathbb{C}^n$, where $n$ is a positive integer, $n\in\mathbb{N}$.
The elements $x\in {\cal H}$ of a complex vector space ${\cal H}$ are called \emph{vectors}.
A \emph{linear morphism} from a complex vector space ${\cal H}_1$ to another complex vector space ${\cal H}_2$ is
a function $f:{\cal H}_1\rightarrow {\cal H}_2$ such that
\begin{enumerate}
\item $f(x+y)=f(x)+f(y),\quad\forall x,y\in {\cal H}_1$.
\item $f(\lambda x)= \lambda f(x),\quad \forall \lambda\in\mathbb{C}, x\in {\cal H}_1$.
\end{enumerate}
A linear morphism $f$ is an \emph{isomorphism} if $f$ is bijective.
A \emph{finite-dimensional} complex vector space ${\cal H}$
is a complex vector space isomorphic to $\mathbb{C}^n$ for some $n\in\mathbb{N}$.
The number $n$ is called the \emph{dimension} of ${\cal H}$. Following Dirac's notation, the pair $({\cal H},\langle -|-\rangle)$ is a finite dimensional \emph{Hilbert space} if ${\cal H}$ is a finite dimensional complex vector space and $\langle -|-\rangle$ is an inner product in ${\cal H}$. An \emph{inner product} in ${\cal H}$ is a function $\langle -|-\rangle: {\cal H} \times {\cal H} \rightarrow \mathbb{C}$ satisfying,

\begin{enumerate}
\item $\langle x|y\rangle=\overline{\langle y|x\rangle},\quad\forall x,y\in {\cal H}$.
\item $\langle \lambda_1 x_1+\lambda_2 x_2|y\rangle = \lambda_1 \langle x_1|y\rangle +\lambda_2 \langle x_2|y\rangle ,\quad \forall \lambda_1,\lambda_2\in\mathbb{C}, x_1,x_2,y\in {\cal H}$.
\item $\langle x|x\rangle >0,\quad\forall x\neq 0, x\in {\cal H}$.
\end{enumerate}
As an example, $\mathbb{C}^n$ with the usual inner product, $\langle v|w\rangle=\sum_{i=1}^n v_i\overline{w_i}$, is a Hilbert space of dimension $n$.
Finally, a set $\{x_1,\dots,x_n\}\subseteq {\cal H}$ in an $n$-dimensional Hilbert space is a \emph{orthonormal basis} if $\langle x_i|x_j\rangle=0$ for all $1\leq i<j\leq n$ and $\langle x_i|x_i\rangle=1$ for all $i=1,\dots,n$.

Now, in order to establish the equivalence between Heisenberg's and Schr\"odinger's theories, von Neumann recalled the isomorphism between the separable Hilbert spaces $\ell^2$ and $L^2(\Omega)$ \cite{VN},
\[
\ell^2=\left\{(x_1,x_2,\dots)\,\colon\,\sum_{i=1}^\infty |x_i|^2<\infty \right\},\quad
L^2(\Omega)=\left\{f:\Omega\to\mathbb{C}\,\colon\, \int_\Omega |f|^2<\infty\right\}/\sim,
\]
where $f\sim g$ if $f=g$ almost everywhere (it is known that any infinite dimensional separable Hilbert space is isomorphic to $\ell^2$). Von Neumann explained that the isomorphism between $\ell^2$ and $L^2(\Omega)$, gave a relation between the functions in the discrete space $\mathbb{N}$ and functions in the continuous state space $\Omega$ of a mechanical system, namely, the wave functions. With this relation in mind and assigning the role of wave functions in $L^2(\Omega)$ to the vectors in $\ell^2$, von Neumann was creating a bridge between the two frameworks. 

Dirac and von Neumann had reformulated the mathematical account of the theory in terms of vector spaces, leaving matrices behind. But since, from a mathematical perspective, vectors could re-generate the whole matrix space, there seemed to be a complete {\it equivalence} with no loss of information. At this point it is vital for our work to stress an essential distinction between the abstract reference to matrices and vectors and their specification relative to a reference frame or basis. While this distinction is irrelevant for mathematicians it is essential for physicists when attempting to represent a {\it state} of a system as an {\it invariant}, as that which remains {\it the same} with respect to different reference frames (or bases). Thus, it is important to distinguish, firstly, between matrices in the canonical basis in $\mathbb{C}^{n\times n}$ and abstract matrices representing linear transformations between abstract vector spaces in $B(\mathcal{H})$. As we mention earlier, while in $\mathbb{C}^{n\times n}$ there exists an array of complex numbers $a_{ij}$ (a table of numbers) such that the matrix is equal to those numbers, in the abstract setting there is no such reference. Clearly, changing the basis, changes the matrix to a \emph{similar} one. Hence, the representation of an abstract matrix in $\mathbb{C}^{n\times n}$ depends on the basis, but the similarity class (e.g., {\it eigenvalues}) remains invariant and independent from the basis. In order to make the distinction clearer, we use the notation $[A]_B$ for a matrix in $\mathbb{C}^n$ represented in the basis $B$ and $A$ for an abstract matrix in $B(\mathcal{H})$ (abstract linear transformations in the finite dimensional vector space $\mathcal{H}$).\footnote{Let us remark that the situation in infinite dimensions is different. In principle, one could define infinite matrices as an infinite array of numbers, but this definition has a lot of difficulties regarding convergence. A different approach is to take the abstract setting and work with bounded operators, linear transformations between abstract Hilbert spaces, see \cite{VN}.} Secondly, let us stress the fact that Dirac's notation of vectors in terms of {\it kets}, $| \psi \rangle$, implies right from the start the reference to a specific basis. Obviously, if $| \psi \rangle = a | \phi_1 \rangle + b | \phi_2 \rangle$, both $| \psi \rangle$ and  $a | \phi_1 \rangle + b | \phi_2 \rangle$ are {\it the same} vector but written in different bases. In order to refer to an {\it abstract vector} we will use the following standard notation $\psi$. This distinction will be essential for sections 3 and 5. The vector $\psi$ can be seen as a rank one {\it abstract matrix} through the following operation $\psi^\dag \psi$. Indeed, if ${\cal H}$ is an abstract $n$-dimensional complex vector space and $B({\cal H})$ is the space of $n\times n$ abstract matrices, then we can relate the space of vectors ${\cal H}$ with the space of rank 1 matrices $B({\cal H})$ through the map:
\[
\nu:{\cal H}\to B({\cal H}),\quad \nu(x):=\psi^\dag \psi.
\]
Let us mention two relevant properties of $\nu$. The first one is that $\nu$ is not {\it surjective}. In fact,  its image is equal to the set of rank one matrices. The second relevant property of $\nu$ is that it is {\it injective}. Hence, we can think of the space of vectors as a subset of the space of matrices. In other words, the vector space is much ``smaller''\footnote{Smaller in the sense that $\nu$ is an injection, and hence, the space of vectors is included in the space of matrices.} than the matrix space (see Appendix C). We will come back to this point later in the final sections of this work.

\section{The (Inconsistent) Bohrian-Positivist Scheme}

The development of QM can be directly related to the empirical-positivist {\it Zeitgeist} which commanded the end of the 19th century and the beginning of 20th. In this context, Ernst Mach is responsible for having shifted the understanding of theories as well as its reference from the description of reality to the algorithmic prediction of empirical observations ---which he called sensations. It was this radical move which, in turn, allowed physicists like Einstein and Heisenberg to develop new theories which departed radically the classical modern representation of physics such as special relativity and QM. In the philosophical arena, Mach's followers congregated in the Vienna circle advanced a logical-linguistic understanding of physical theories that would restrict their reference to observations ---understood as unproblematic {\it givens} of experience--- grounding a common foundation for the positivist program of a unified science \cite{VC}. Essential to the positivist project was the rejection of all forms of metaphysics ---understood by them as an ungrounded narrative about ``unobservable'' phenomena. The proposal was to abandon the modern metaphysical claim according to which ``theories attempt to represent reality'' and replace it by a more sober and skeptic account of them as  ``economies of experience''. However, instead of leaving behind the modern substantialist account of physical reality ---which Mach himself had strongly criticized (see \cite[pp. 78-79]{Mach70})--- logical positivism would retain this metaphysical discourse first as a ``natural'' or ``commonsensical'' foundation required to account for experience ---something explicit in the work of Rudolph Carnap who embraced what he called a ``physicalist language'' as a ``natural'' way to refer to ``things within space and time'' \cite{Carnap28}--- and then, after Neurath's critic to {\it protocol statements}, as a completely ungrounded, always changing, ``way of talking'' about subjective observations. It is in this new anti-realist context that Niels Bohr would create in 1913 a successful quantum model in which a series of ``magical'' inconsistent rules capable of predicting spectral lines in the lab would become ---also inconsistently--- supplemented with an ``atomist-planetary'' narrative. Bohr claimed that his model represented the Hydrogen atom as a ``small planetary system'' in which orbits were discretely quantized and electrons could ``jump'' from one orbit to the next in a non-causal fashion. Even though the picture had no theoretical nor empirical support, it captivated most physicists who were eager to restore their substantialist belief in a world composed of particles. Bohr's inconsistent model would be then embraced not only by physicists but also by logical positivists. For after all, the model did provide the correct predictions of intensive patterns, and that, according to positivists, was all that mattered. Thus, while physicists seemed to believe in the real existence of such tiny planetary systems, positivists would tolerate Bohr's narrative as ``a way of talking'' about ``useful fictions'' (see \cite{deRondeFM21, Faye21}). The Danish physicist was awarded the Nobel prize in 1922 and crowned as the ruler of the atomic Kingdom. As we shall discuss in this work, Bohr's effective strategy to introduce fictional narratives in order to bridge the many gaps of a self-contradictory story would be repeatedly applied in the context of QM. 

Even though their presentation was quite different, there is a series of common presuppositions that can be found in both the works of Bohr and the followers of Mach (see for a detailed analysis of their common approach \cite{Faye21}). The most important point of agreement is their reference to classical experience as a ``common sense'' standpoint for the development of theories. While the Danish physicist would stress the need to consider classical theories in order to discuss about phenomena, logical positivists would stress the reference to that which is naturally {\it given} within experience ``when we open our eyes''. As they would write in their famous Manifesto \cite{VC}: ``Everything is accessible to man; and man is the measure of all things. Here is an affinity with the Sophists, not with the Platonists; with the Epicureans, not with the Pythagoreans; with all those who stand for earthly being and the here and now.'' Also, while logical positivists would stress the need to erase completely metaphysical discourse from physics, Bohr would constantly worn his followers against the possibility to develop new concepts in order to describe the quantum realm. According to Bohr \cite[p. 7]{WZ}: ``[...] the unambiguous interpretation  of any measurement must be essentially framed in terms of classical physical theories, and we may say that in this sense the language of Newton and Maxwell will remain the language of physicists for all time.'' More importantly, ``it would be a misconception to believe that the difficulties of the atomic theory may be evaded by eventually replacing the concepts of classical physics by new conceptual forms.''  However, there was also an essential difference between them, while positivists would despise all forms of metaphysics Bohr would read the limit imposed by QM to the possibility of physical representation of reality not as an epistemological limit but as an ontological finding about reality-in-itself, namely, its ontological irrepresentability. His whole scheme would then become grounded in two principles. While {\it correspondence} would allow Bohr to impose an irrepresentable yet existent ``limit'' between the classical and quantum theories that would become essential in order to justify his interpretation of QM as ``a rational generalization of classical mechanics'' \cite{Bokulich05}, {\it complementarity} would allow him to escape the objective Kantian reference to categorically defined `moments of unity' replacing it by the intersubjective reference to measurement outcomes (essentially detached from representation). In turn, these two principles would be explicitly followed by Dirac and von Neumann in order to reach a ``Standard'' formulation of QM (SQM) during the early 1930s \cite{deRonde22, deRondeFM21} ---also known by physicists under different names such as ``the Copenhagen interpretation'', the ``Dirac-von Neumann formulation'' or, simply put, ``non-relativistic QM''. As we shall see in the following sections, the general inconsistencies present within the Bohrian-positivist scheme for understanding physical theories would become extended in Dirac's and von Neumann's formulation of QM through the re-definition of the concepts of (quantum) {\it state} and {\it mixture}. 


At this point it is important to remark that SQM has remained essentially untouched since its original formulation and even the notation proposed by Dirac has remained the orthodox way of expressing QM within mainstream physical journals. Of course, none of this has anything to do with the interpretational debate that populates philosophical journals and has remained essentially unnoticed by physicists.\footnote{As Maximilian Schlosshauer \cite[p. 59]{Schlosshauer11} has recently described: ``It is no secret that a shut-up-and-calculate mentality pervades classrooms everywhere. How many physics students will ever hear their professor mention that there's such a queer thing as different interpretations of the very theory they're learning about? I have no representative data to answer this question, but I suspect the percentage of such students would hardly exceed the single-digit range.''} This philosophical debate has not only introduced new narratives in terms of propensities, many worlds, flashes, etc. but has even changed the mathematical formalism of the theory, like in the case of Bohmian mechanics and GRW, leading to completely different schemes to that of SQM.

\section{Dirac's (Inconsistent) Definition of (Quantum) State}

The notion of {\it state} in classical physics was of course related to to the general realist aim of the discipline as making possible the reference to a (real) state of affairs in objective-invariant terms independently of particular reference frames or subjective perspectives. As we discussed extensively in \cite{deRondeMassri21} the notion of {\it system} provides an abstract account of the formal-conceptual {\it moment of unity} (e.g., particle, wave) required by physical theories in order to unify the multiplicity found within experience. The explicit definition of a {\it moment of unity} in mathematical and conceptual terms is of course a precondition in order to discuss about the evolution and dynamics of a state of affairs. Without the reference to `something' (which evolves) it becomes simply impossible to discuss in dynamical terms. In this respect, the physical notion of {\it state} has the main purpose to address not only the specific values of the properties of a system within a given specific situation (e.g., the particle 1 has position $p_1$ and velocity  $v_1$ while the particle 2 has position $p_2$ and velocity  $v_2$, etc.), but also the way in which different perspectives can be brought into unity.  This is where the notion of reference frame enters the scene. Indeed, in physics, the specification of the state of each system the theory talks about is always {\it relative} to a {\it reference frame}. The specific values of the properties of a system in a given situation ---namely, the state of the system--- have meaning only when considered as relative to a specific viewpoint. Without a reference frame there can be no consistent definition of the state of a system. In this context, the notion of {\it physical invariance} becomes essential in order to secure the consistency of the different reference-frame dependent representations of the (same) state of affairs. In conclusion, it is the consistent translation between the different reference frame dependent representations which allows us to talk about {\it the same} state of affairs as something completely detached from a particular viewpoint.

Contrary to this understanding of physical theories as making reference to a detached state of affairs, in his 1930 book {\it The Principles of Quantum Theory} ---following the Bohrian-positivist premises--- Dirac would impose a radically new understanding of the meaning of {\it state} ---which has remained untouched up to the present. Each (quantum) {\it state}, mathematically represented by ---what he himself called--- a {\it ket}, $|x\rangle$, would be linked to a single measurement observation; i.e., a `click' in a detector. As we already mentioned, Dirac's notation imposed an explicit link between a {\it ket} and a specific reference frame (or basis). This means that a {\it ket} is not an {\it abstract vector}, $x$, but one already represented in a specific basis. Explaining his approach, Dirac would argue that {\it the same system} could be represented in terms of {\it different states}, relative to the different bases: 
\begin{quotation}
\noindent {\small ``[E]ach state of a dynamical system at a particular time corresponds to a ket vector, the correspondence being such that if a state results from the superposition of certain other states, its corresponding ket vector is expressible linearity in terms of the corresponding ket vectors of the other states, and conversely. Thus the state $R$ results from a superposition of the states $A$ and $B$ when the corresponding ket vectors are connected by $ | R \rangle\  = c_1 | A \rangle\ + c_2 | B \rangle\ $.'' \cite[p. 16]{Dirac74}}
\end{quotation}
Thus, depending on the chosen basis (or reference frame), the representation of the system would be given in terms of different states. This means that in Dirac's formulation bases did not relate to {\it the same state}, but instead would impose the distinction between {\it different states}. Each reference frame (or basis) would make reference to a particular state ---in general--- composed of many (superposed) states. The inconsistency ---to which we will come back in section 5--- should already be clear to the attentive reader (see for a more detailed analysis \cite{deRondeMassri21}). However, let us make even more explicit the contradiction that is reached when applying these inconsistent definitions. 

\medskip
\begin{center}
\begin{tabular}{c|l}
Mathematical symbol & Physical interpretation \\ 
\hline
$\Psi$              & Abstract vector state \\
$|\uparrow_x\rangle$                    &  A one term state representing $\Psi$ in the basis $\{ |\uparrow_x\rangle, |\downarrow_x\rangle\}$. \\ 
$a|\uparrow_y\rangle+b|\downarrow_y\rangle$ &  A superposition state as a sum of different states \\ & when representing $\Psi$ in the basis $\{ |\uparrow_y\rangle, |\downarrow_y\rangle\}$.\\
\end{tabular}
\captionof{table}{Mathematical and physical accounts of vectors and states.}
\end{center}
\medskip
From a mathematical perspective it is clear that the first column are different representations of the same state vector. Yet, from a physical perspective, things are different. Even though there is no contradiction in mathematical terms, from a physical perspective {\bf a state that is certain cannot be uncertain}. The state $|\uparrow_x\rangle$ is {\it certain} when measured in the basis $\{ |\uparrow_x\rangle, |\downarrow_x\rangle\}$, but {\it the same} abstract vector (i.e., the same state) is also given by the state $a|\uparrow_y\rangle+b|\downarrow_y\rangle$ which is {\it uncertain} when measured in the basis $\{ |\uparrow_y\rangle, |\downarrow_y\rangle\}$. This is an inconsistency, the same state cannot be certain and uncertain. The inconsistency appears because {\it the same} abstract state vector $\Psi$ is related through the different basis dependent representations of the vector (i.e., $|\uparrow_x\rangle$  and  $a|\uparrow_y\rangle+b|\downarrow_y\rangle$) to different physical states of affairs. A mathematical equivalence is not a necessarily a physical equivalence. The term {\it state} is then used to refer to different incompatible situations which are then also considered to be the same in abstract mathematical terms. The states $|\uparrow_x\rangle$,  $|\uparrow_y\rangle$ $|,\downarrow_y\rangle$ are different states and yet are also the same state. This confusion, created by Bohr and Dirac, has remained part of the mainstream physics literature about SQM. 

The radical transformation in the meaning of (quantum) state imposed by Dirac can be only understood in the context of the anti-realist 20th century re-foundation of physics commanded by Niels Bohr and logical positivists. Following Bohr, the classical presupposition of physics according to which a system can only have a single state at a time would be abandoned as a prejudice of the past and replaced by a new contextual understanding of (quantum) states determined by the choice of the experimental apparatus (linked to a specific basis). But since the metaphysical representation of reality had been already abandoned, the lack of an objective system was not regarded as a problem. After all, as Dirac \cite[p. 10]{Dirac74} would argue, ``the main object of physical science is not the provision of pictures, but the formulation of laws governing phenomena and the application of these laws to the discovery of phenomena. If a picture exists, so much the better; but whether a picture exists of not is a matter of only secondary importance.'' In short, it is ``important to remember that science is concerned only with observable things''. So far so good. But there was still an obstacle. Dirac's choice to restrict {\it observational certainty} to a binary representation had been dogmatically exported from an implicit reference to metaphysical atomism where, in fact, the representation of physical reality had been conceived in terms of binary valued properties ---i.e., an {\it actual state of affairs} (see \cite[Definition 1.1]{deRondeMassri18a}). According to Dirac each and every `click' in a detector implied the observation of the state of a particle ---even though there was no representation of the particle itself. Thus, a double inconsistent reference was created in relation to the state of affairs. On the one hand a state was linked to each single observation, but on the other a state of affairs was also described in terms of a quantum superposition, namely, a linear combination of different states. Since the conceptual account of the (quantum) state of affairs remained completely undefined ---considered as irrepresentable--- the discussion was cleverly re-directed towards the debate about the ---also irrepresentable--- measurement processes which linked superpositions with (binary) outcomes. This strategy had been already applied by Bohr in his famous model of the atom and also in his debates with Einstein ---including of course his reply to the EPR paper \cite{Bohr35}. As Bohr had done before, Dirac simply blamed his controversial (inconsistent) definition of quantum state on the limits of QM to represent microscopic entities. The problem, given that physics ---according to positivists--- had to be understood as predicting observations, became then the following: How can a quantum superposition provide certain (binary) knowledge about measurement observations? It is at this point that Dirac was forced to introduce a new ``quantum jump'' that would allow to bridge the gap between quantum superpositions and the single (binary) observations he wanted to refer to: 
\begin{quotation}
\noindent {\small ``When we make the photon meet a tourmaline crystal, we are subjecting it to an observation. We are observing whether it is polarized parallel or perpendicular to the optic axis. The effect of making this observation is to force the photon entirely into the state of parallel or entirely into the state of perpendicular polarization. It has to make a sudden jump from being partly in each of these two states to being entirely in one or other of them.'' \cite[p. 7]{Dirac74}}
\end{quotation}
As he had already learned from Bohr, the irrepresentability of this new quantum jump would be easily blamed to the theory itself. The mantra was simple: ``It is weird because it is quantum!'' 

To sum up, Dirac's work implies a replacement of the modern formal-conceptual definition of {\it state} in terms of {\it invariance} and {\it objectivity} by a double inconsistent reference, on the one hand, to abstract vectors and, on the other, to the empirical observation of (binary) measurement outcomes. All of this supplemented by an inconsistent atomist narrative about strange quantum particles with no definite valued properties that could not only behave like waves or corpuscles (depending on the context of measurement) but could also evolve through ``quantum jumps'' in ways that could not be theoretically described. Dirac's scheme can be then understood as grounded in three essential points: 
\begin{enumerate}
\item[I.] {\it The vectorial restriction of the matrix space to rank 1 matrices} (see Appendix B).
\item[II.] {\it The re-definition of the notion of state in terms of observational binary certainty.} 
\item[III.] {\it The ad hoc introduction of the collapse (or projection) postulate as a consequence of measurement.} 
\end{enumerate}
The inconsistency of this scheme, which has remained either unnoticed or tolerated in the physics and philosophy communities, is also linked to the distinction between pure states and mixtures to which we will now turn our attention.

\section{Landau and von Neumann's Purity-Mixture Distinction}

As we discussed above, Heisenberg's matrix mechanics was almost immediately replaced by Sch\"odinger's wave mechanics and supplemented by Born's probabilistic interpretation of the quantum wave function. While the wave formulation re-introduced ---at least apparently--- the continuum promising to restore a spatiotemporal representation, Born's rule re-introduced not only an ignorance interpretation about measurement outcomes but also a discursive reference to ``particles''. Once again, the many gaps were blamed on the theory itself and its essential ---yet irrepresentable--- randomness. As described by von Neumann:
\begin{quotation}
\noindent {\small ``This concept of quantum mechanics, which accepts its statistical expression as the actual form of the laws of nature, and which abandons the principle of causality, is the so-called `statistical interpretation'. It is due to M. Born, and is the only consistently enforceable interpretation of quantum mechanics today ---i.e., of the sum of our experience relative to elementary processes.'' \cite[p. 136]{VN}} 
\end{quotation}
Of course, the fact that QM implied a non-Kolmogorovian probability measure which precluded an ignorance interpretation about an underlying actual state of affairs was already known to everyone. The probability measure was given by Gleason's theorem:
\begin{theorem}
Let $\mu$ be a measure on the set of orthogonal projections of a separable Hilbert space of dimension at least three. There exists a positive semi-definite self-adjoint operator $\rho$ of the trace class such that for all orthogonal projection $P$, 
\[
\mu(P) = \Tr(\rho P).
\]
Recall that a \emph{measure} $\mu$ is a function to $[0,1]$
such that if $\{P_i\}$ is a countable collection of mutually commuting projections, then $\mu(\sum P_i)=\sum\mu(P_i)$.
\end{theorem}
\noindent As repeatedly remarked by Einstein and Schr\"odinger, the problem was not ---as argued by Bohr, Born, Dirac and von Neumann--- related to the measurement process, it was already present in the definition of the {\it quantum state} even before the process of collision or measurement could even be considered. As Schr\"odinger would criticize the notion of quantum state in his famous ``cat paper'': 
\begin{quotation}
\noindent {\small ``One should note that there was no question of any time-dependent changes. It would be of no help to permit the [quantum mechanical] model to vary quite `unclassically,' perhaps to `jump.' Already for the single instant things go wrong. [...] if I wish to ascribe to the model at each moment a definite (merely not exactly known to me) state, or (which is the same) to {\it all} determining parts definite (merely not exactly known to me) numerical values, then there is no supposition as to these numerical values {\it to be imagined} that would not conflict with some portion of quantum theoretical assertions.'' \cite[p. 156]{Schr35}}
\end{quotation}
On the contrary, the Bohrian solution was grounded on the claim that QM had reached the limits of representation of reality-in-itself. Quantum particles were unknowable, irrepresentable, due to the uncontrollable interaction, produced by the {\it quantum of action}, between quantum and classical systems ---a process which was also irrepresentable. The problem of adequately defining the reference of the theory was then replaced by the problem of accepting the limits of representation. And it is in this context that it was then possible for Born to introduce a new quantum jump that would bridge the gap between quantum superpositions and measurement outcomes. The ``collapse'', after being explicitly stated by Dirac, would be also reinforced by von Neumann: 
\begin{quotation} 
\noindent {\small ``if the system is initially found in a state in which the values of R cannot be predicted with certainty, then this state is transformed by the measurement M of R (in the example above, M1) into another state: namely, into one in which the value of R is uniquely determined. Moreover, the new state, into which M places the system, depends not only upon the arrangement of M but also upon the result of the measurement M (which could not be predicted causally in the original state), because the value of R in this new state must actually be equal to this M-result.'' \cite[p. 138-139]{VN}} \end{quotation}

Bohr, Born and Dirac were introducing ---irrepresentable--- ``quantum jumps'' in order to ``explain'' the evolution of the ---also irrepresentable--- ``quantum particles''. It is this move which allowed them to reintroduce an ignorance interpretation of quantum probability referring not to an actual state of affairs but instead, to binary observable events (or quantum states). It is in this conflictive context that two mathematicians, the Russian Lev Landau and the Hungarian John von Neumann, would develop a statistical account of the theory of quanta where the notion of mixture ---exported from classical statistical theories--- would come to play an essential role. As Eugene Wigner would recall: 
\begin{quotation}
\noindent {\small ``The concept of the mixture was put forward first by L. Landau, Z. Physik, 45, 430 (1927). A more elaborate discussion is found in J. von Neumann's book (footnote 4), Chapter IV. A more concise and elementary discussion of the concept of mixture and its characterisation by a statistical (density) matrix is given in L. Landau and E. Lifshitz, Quantum Mechanics (London: Pergamon Press, 1958), pp. 35-38.'' \cite[p. 183]{Wigner95}}
\end{quotation}
But while the motivation that inspired Landau was linked to the need of describing the subsystems of a composite quantum system by a state vector \cite{Landau27}, von Neumann would introduce the notion of mixture in order to develop what he called ``quantum statistical mechanics'' as well as ---following Bohr--- a new theory of quantum measurements. L\'eon van Hove explains:
\begin{quotation}
\noindent {\small ``The analysis, carried out in a paper of 1927\footnote{J. von Neumann, ``Wahrscheinlichkeitstheoretischer Aufbau der Quantenmechanik'', {\it Nachr.
Ges. Wiss.} G\"ottingen (1927) pp. 245-272.}, introduced the concept of statistical matrix for the description of an ensemble of systems which are not necessarily all in the same quantum state. The statistical matrix (now often called $\rho$-matrix although von Neumann's notation was $U$) has become one of the major tools of quantum statistics and it is through this contribution that von Neumann's name became familiar to even the least mathematically minded physicists.'' \cite[p. viii]{VN}}
\end{quotation}


On the one hand, following the idea that QM made reference to particles, Landau wrote in \cite{Landau27} about \emph{coupled systems} and the necessity of a probability ensemble. He pointed out that if a system is coupled with another, then the uncertainty grows. If the first system can be described by quantities $\{a_i\}$, 
\[
|x\rangle = \sum_i a_i|x_i\rangle
\]
and the second system by $\{b_j\}$, 
\[
|x'\rangle = \sum_j b_j|x'_j\rangle,
\]
then the two system together are described by $\{c_{ij}=a_ib_j\}$,
\[
|xx'\rangle = \sum_{ij} a_ib_j |x_ix_j'\rangle.
\]
If the two systems are coupled, the coefficients $c_{ij}$ depend on time and can no longer be resolved as a product. On the other hand, in \cite[\S III]{VN}, von Neumann would reduce all assertions of QM to the Born Rule, $\langle R\phi,\phi\rangle$, which represents the expectation value of the quantity $R$ in the state $\phi$ (here $\phi$ is a vector). In Chapter IV, he analyzes the situation of a mixture of states, $\{(\phi_i,p_i)\}$, and he writes that this situation occurs when we do not know what state is actually present. After some mathematical arguments he concludes that the expectation  value of the quantity $R$ in this situation is also the Born Rule, $\Tr(UR)$, where $U=\sum p_i\phi^\dag\phi$. 

To sum up, even though it was recognized that quantum probability could not be understood in terms of epistemic ignorance, an ignorance interpretation was anyhow applied in QM through the artificial introduction of the purity-mixture distinction. This inconsistent reintroduction of ignorance had two different layers. While in the first layer Born had reintroduced ignorance through a probabilistic account of the quantum wave function related to the finding of a single measurement outcome (i.e., a {\it pure state}), the second layer introduced by Lev Landau and John von Neumann would reintroduce matrices, not as part of the mathematical formalism itself, but as a classical statistical measure of ignorance of the first  layer of purity (see Appendix B). As we shall see in the following sections both notions of purity and mixture are simply inconsistent from a formal, conceptual and operational perspective.

\section{The Inconsistency of Pure States}

Dirac would re-define of the notion {\it state} in terms of {\it abstract vectors} on the one hand, and in terms of {\it basis represented vectors} (i.e., {\it kets}) on the other. This deep transformation of the meaning of state in QM was criticized by both Pauli and Schr\"odinger. Arthur Fine \cite[p. 94]{Schlosshauer11} comments: ``Wolfgang Pauli thought that using the word `state' ({\it Zustand}) in QM was not a good idea, since it conveyed misleading expectations from classical dynamics.'' Erwin Schr\"odinger \cite[p. 153]{Schr35} would also criticize Dirac's use of the notion of state arguing that: ``The classical concept of {\it state} becomes lost [in QM], in that at most a well-chosen {\it half} of a complete set of variables can be assigned definite numerical values.'' As discussed in detail in \cite{deRondeMassri21}, the essential inconsistency present in Dirac's re-definition of (quantum) state is related to the misuse of {\it reference frames} as a precondition to account for any operational content within the same physical situation. As remarked by Einstein: 
\begin{quotation} \noindent {\small ``The concept does not exist for the physicist until he has the possibility of discovering whether or not it is fulfilled in an actual case. We thus require a definition of [the concept] such that this definition supplies us with the method by means of which, in the present case, he can decide by experiment whether or not [the concept] occurred.'' \cite[p. 26]{Einstein20}} \end{quotation}
 
Let us discuss this in some detail. The operational definition of the notion of pure state commonly applied in the orthodox literature runs as follows: if a quantum system is prepared in a {\it maximal basis} so that there is a {\it maximal test} yielding with certainty a particular outcome, then it is said that the quantum system is in a \emph{pure state}. Unfortunately, this operational definition of state is linked to a  reference frame (or basis), and consequently is not invariant. Any basis containing --in Dirac's notation-- the {\it ket} $| x \rangle$ will give binary certainty when measuring in the direction $x$ that state. 
\begin{definition}[Operational Purity] Given a quantum system in the state $|\psi \rangle$, there exists an experimental situation linked to that basis (in which the vector is written as a single term) in which the test of it will yield with certainty (probability = 1) its related outcome. 
\end{definition}
\noindent Von Neumann's application of this notion in the context of quantum logic is also explicit as related to his definition of {\it actual property}, something applied in the many operational approaches that were developed during the 1960s and 1970s (see for a detailed analysis \cite{dDFInternet}). In short, a property is {\it actual} if given a specific experimental set up we know with certainty (probability = 1) the result of the future outcome (see also \cite{Aerts81, Piron76}). 

However, there is also a reference to pure states in basis-independent terms, for any rotation of a pure state ---now understood as a vector in purely abstract terms; i.e., independent of any basis--- is also considered to be {\it the same} pure state. This is justified in terms of the mathematical fact that an abstract vector is defined as an invariant element under rotations.  
\begin{definition}[Abstract Purity]\label{pure} An abstract unit vector (with no reference to any basis) in Hilbert space, $\Psi$, is a pure state. In terms of density operators $\rho$ is a pure state if it is a projector, namely, if Tr$(\rho^2) = 1$ or $\rho = \rho^2$. 
\end{definition}
\noindent It is at this point that we need to clearly distinguish between a purely abstract vector $\psi$ and its specific representations in different bases, say in $|\psi\rangle$ or in $c_1 |\phi_1\rangle + c_2 |\phi_2\rangle$.\footnote{As shown in detail in \cite[Sect. 4]{deRondeMassri19} this distinction becomes explicitly visualizable through the use of graph theory (see figures 1 and 6 of the mentioned reference).} As discussed in detail \cite{deRondeMassri21}, there is an essential equivocity within these two (inconsistent) definitions of purity. While the operational purity is basis-dependent (i.e., it explicitly depends on the {\it maximal basis}) and consequently non-invariant, vectorial purity provides an invariant definition but has no operational content whatsoever. Thus, while the operational definition destroys the invariant reference of the mathematical formalism, the latter abstract definition lacks an experimental counterpart. Furthermore, there is no equivalence between these two distinct definitions. While operational purity implies vectorial purity the converse is false. An abstract vector does not imply a specific basis dependent representation of it. As we will discuss in the following section this inconsistency present in the notion of {\it pure state} is ---obviously--- extended to the notion of {\it mixture} which is explicitly grounded on it. 

It is important to stress that the definitions we have discussed above are commonly presented in advanced textbooks and applied within the mainstream literature. Just to provide a few examples where the reader can find these definitions explicitly we mention the famous books by Nielsen and Chuang, {\it Quantum Computation and Quantum Information} \cite{NielsenChuang10}, the one by Bengtsson and Zyczkowski, {\it Geometry of Quantum States} \cite{BZ17}, the Preskill Lecture notes on quantum information \cite{Preskill97}, Mermin's book {\it Quantum Computer Science} \cite{Mermin07} and {\it Quantum Computing since Democritus} by Scott Aaronson \cite{Aaronson13}.

\section{The Inconsistency of Mixed States}

As discussed in section 4, Landau and von Neumann introduced, for different reasons, the classical statistical notion of mixture within the theory of quanta during the late 1920s. Ever since, this notion has become an essential element within SQM and is  presented in textbooks in the following manner. Let $\mathcal{H}$ be a Hilbert space, a density operator $\rho$  (i.e., a positive trace class  operator with trace 1) is called a \emph{state}. Being positive and self-adjoint, the eigenvalues of $\rho$ are non-negative and real and it is possible to diagonalize it. Pure states are represented by rank 1 matrices which in their diagonal form will be given by $(1,0,\dots,0)$ ---where the place occupied by the `1' is unimportant. In this case,  $\rho$ is equal to  $vv^{\dag}$ for some normalized abstract vector $v\in\mathcal{H}$. However, if the rank of $\rho$ is grater than 1 (or equivalently if $\mbox{Tr}(\rho^2)<1$), the state is called  a \emph{mixed state}, or in short a {\it mixture}.\footnote{As remarked by Ugo Fano \cite[p. 75]{Fano57}: ``The name `statistical matrix' is often used instead of `density matrix'. The name density matrix itself relates to the correspondence between $p$ and the distribution function $\rho(q_i, p_i)$ in the phase space of classical statistical mechanics. This correspondence has been developed by Wigner \cite[Sect. 9]{Wigner32}.''} The distinction between pure and mixed states is then understood, as  David Mermin \cite[p. 754]{Mermin98b} explains, in terms of ignorance and certainty:  ``[...] people distinguish between pure and mixed states. It is often said that a system is in a pure state if we have maximum knowledge of the system, while it is in a mixed state if our knowledge of the system is incomplete.'' As remarked by Nancy Cartwright \cite{Cartwright72}:  ``The ignorance interpretation is the orthodox interpretation for mixtures. [It] asserts that each member of the collection is in one of the pure states in the sum ---it is only our ignorance which prevents us from telling the right pure state for any specific member.'' However, such introduction of an ignorance interpretation of matrices, in analogous terms to classical statistical theory, is clearly problematic. This was already clear to Schr\"odinger who in a letter to Einstein world write the following:
\begin{quotation}
\noindent\small{``It seems to me that the concept of probability is terribly mishandled these days. Probability surely has as its substance a statement as to whether something \emph{is} or  \emph{is not} the case ---an uncertain statement, to be sure. But nevertheless it has meaning only if one is indeed convinced that the something in question quite definitely \emph{is} or  \emph{is not} the case. A probabilistic assertion presupposes the full reality of its subject.'' \cite[p. 115]{Bub97}}
\end{quotation}
Indeed, the applied ignorance interpretation of mixtures refers to {\it pure states} which do not provide a consistent representation of ``quantum particles''. As Schr\"{o}dinger would remark \cite[p. 188]{Schr50}: ``We have taken over from previous theory the idea of a particle and all the technical language concerning it. This idea is inadequate. It constantly drives our mind to ask information which has obviously no significance.'' These remarks show the difficulties of introducing in a meaningful fashion the notion of mixture also discussed explicitly by Schr\"odinger in \cite{Schr36}. However, to use a phrase by Park \cite[p. 225]{Park88}, he ``had exposed only the tip of the iceberg.'' Indeed, there are much deeper problems with the notion or mixture.    

Let us first recall that Heisenberg's original formulation of QM makes no distinction whatsoever between rank 1 matrices (which are the only matrices that can be related to vectors) and matrices of rank $\neq 1$. While in the original QM formulation each and every matrix ---independently of the rank--- in each and every basis provides meaningful operational information (given by the list intensive values found within a measurement situation) about a state of affairs (see Appendix A), in Dirac-von Neumann vectorial formulation the only matrices considered as describing quantum states are those of rank 1. The remaining matrix space (i.e., the matrices of rank $\neq 1$) is simply erased (see Appendix B). As a consequence there is, obviously, a huge loss of operational information which, in fact, is the main reason for the re-introduction of these ``erased matrices'' (matrices of rank $\neq 1$) but now re-interpreted as statistical mixtures of pure states (see Appendix C). But also, due to a deep misunderstanding of the meaning and role of {\it reference frames} in physical theories, there is an inconsistent scrambling of two already inconsistent definitions, namely, the operational (basis dependent) definition of purity (i.e., definition 4.1) and the abstract basis independent definition of purity (i.e., definition 4.2). This inconsistent mix might be already clear to the attentive reader, but let us anyhow expose it from a different viewpoint. 

We can understand what is going on through an analysis of the relation between vectors and matrices. The assignment from norm 1 abstract vectors to abstract density matrices can be formalized by using the Veronese embedding (a classical map in algebraic geometry),
\[ 
\nu_2:\mathcal{B}_n\to\mathcal{D}_n,\quad \nu_2(v):=v^\dag v, 
\]
where $\mathcal{B}_n$ is the set of norm 1 vectors and $\mathcal{D}_n$ the set of density matrices,
\[
\mathcal{B}_n:=\{v\in\mathcal{H} \,\colon\,\langle v|v\rangle=1\},\quad
\mathcal{D}_n:=\{\rho\in B(\mathcal{H})\,\colon\,\rho^\dag=\rho,\,\rho\ge 0,\,\Tr(\rho)=1\}.
\]
As we mentioned before, the image of $\nu_2$ are rank 1 abstract density matrices, whose dimension (as a real variety) is $2n-2$, but the dimension of $\mathcal{D}_n$ is quadratic in $n$ (equal to $n^2-1$).  Hence, the variety of rank 1 density matrices  is a low-dimensional subvariety of $\mathcal{D}_n$.  For $n=2$, the image of $\nu_2$ is homeomorphic to the real 2-dimensional sphere, called the Bloch sphere. The interior of the Bloch sphere is a 3-dimensional variety representing the space of maximal rank (i.e., 2) density matrices. For $n=3$, rank 1 density matrices is a space of dimension 4 inside the 8-dimensional space of density matrices.  For $n=10,100,1000$ we have the following table,
\begin{center}
\begin{tabular}{|c||c|c|}
\hline
$n$&$2n-2$&$n^2-1$\\
\hline
\hline
10&18&99\\
\hline
100&198&9999\\
\hline
1000&1998&999999\\
\hline
\end{tabular}
\smallskip
\fig: Dimension of rank 1 density
matrices vs. dimension of $\mathcal{D}_n$.
\end{center}
From this table we can see that the dimension of rank 1 matrices is small in comparison to the dimension of the space of density matrices. Notice that this comparison does not involve bases for we are addressing abstract matrices. 

We can define the map $Eig$ and the following subset of the $n$-simplex,
\[
\Delta_n^{\downarrow}:=\{ (\lambda_1,\dots,\lambda_n)\in\mathbb{R}^n\,\colon\, 
\lambda_1\ge\dots\ge\lambda_n\ge 0,\,\lambda_1+\dots+\lambda_n=1\}.
\]
Notice that we can assign to a density matrix $\rho$ a point
in $\Delta_n^{\downarrow}$ given by its eigenvalues. This assignment 
defines a function
\[
Eig:\mathcal{D}_n\to\Delta_n^{\downarrow}
\]
which is invariant under the unitary group, in other words, the eigenvalues of $\rho$ are equal to the eigenvalues of $U\rho U^{\dag}$,
\[
Eig(\rho)=Eig(U\rho U^{\dag}),\quad \forall U\in \mathcal{U}_n.
\]
An important remark is that, following its mathematical definition, abstract pure states (density matrices of rank 1) are mapped to the point $(1,0,\dots,0)$. In fact, density matrices of rank $<n$ are mapped to a set of measure zero. This implies that a generic density matrix has maximal rank. Summarizing, $Eig(\rho)$ is the set of eigenvalues of $\rho$ sorted in  non-increasing order. For example, if $\rho\in\mathcal{D}_3$,  then $Eig(\rho)=(\lambda_1,\lambda_2,\lambda_3)$ where 
$\lambda_1\ge\lambda_2\ge\lambda_3\ge 0$ and $\lambda_1+\lambda_2+\lambda_3=1$. Hence, if $\mathrm{rk}(\rho)=1$,  $Eig(\rho)=(1,0,0)$ and if $\mathrm{rk}(\rho)=2$, 
then $Eig(\rho)=(\lambda,1-\lambda,0)$, where $1/2<\lambda<1$.
All this information can be visualized using a triangle (or a tetrahedron
in dimension 4 or a $n$-simplex in dimension $n$). In purely abstract terms (i.e., without any reference to bases) the vertices of the triangle correspond to rank 1 density matrices, the sides to rank 2 density matrices and the interior to rank 3 density matrices,
\begin{center}
\begin{tabular}{ccc}
\includegraphics[width=8em]{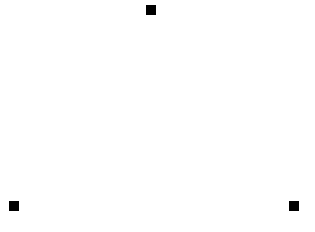}&
\includegraphics[width=8em]{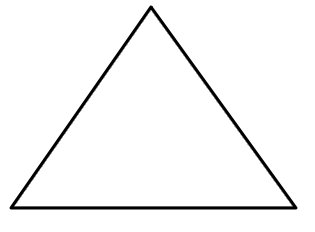}&
\includegraphics[width=8em]{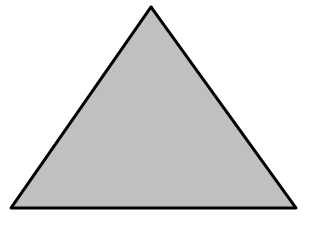}\\
\hfill\fig: Rank 1&\hfill\fig: Rank 2&\hfill\fig: Rank 3
\end{tabular}
\end{center}

\noindent In dimension 4, the density matrices of maximal rank are mapped to the interior of the tetrahedron and the density matrices of rank 1 to the vertexes.  
\begin{center}
\includegraphics[width=8em]{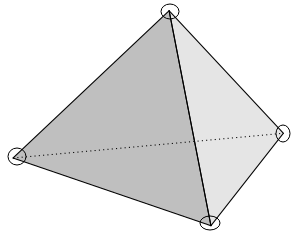}

\hfill\fig: Rank 1 vs. Rank 4
\end{center}

\noindent As we already mentioned, these points do not consider the specification of any basis. This means we are only considering pure states in terms of abstract vectors (definition 4.2). It is from this standpoint that the whole abstract matrix space can be reconstructed from a geometrical viewpoint. Instead of visualizing the eigenvalues of the matrices we can give a geometric description inside $\mathcal{D}_n$ by using the notion of convex join and the Veronese map defined before,
\[
\nu_2:\mathcal{B}_n\to\mathcal{D}_n.
\]
Assume $n>1$. The \emph{convex join} between two sets, $X$ and $Y$, denoted $\mathrm{Conv}(X,Y)$, is defined as all the convex combinations between the elements of $X$ and $Y$,
\[
\mathrm{Conv}(X,Y):=\{ tx+(1-t)y\,\colon\,
x\in X,\,y\in Y,\,0\le t\le 1\}.
\]
Let us denote by 
$R_1:=\mathrm{Im}(\nu_2)=\{v^\dag v\,\colon\,v\in\mathcal{B}_n\}$ 
to the image of $\nu_2$.
As we already know, $R_1$ is the set of all rank 1 density matrices
and let $R_k$ be defined inductively as $R_k:=\mathrm{Conv}(R_1,R_{k-1})$. Now,
it is easy to prove that $R_k$ is the set of rank $\le k$ density matrices
and then, 
any matrix in $R_k$ can be written as a convex combination
of $k$ rank 1 density matrices, $1\le k\le n$.
In particular, any density matrix is a convex combination of rank 1 density matrices, 
that is, $R_n=\mathcal{D}_n$. In other words, the convex hull of $R_1$ 
is exactly $\mathcal{D}_n$.
Then, the space $\mathcal{D}_n$, which has real dimension $n^2-1$, contains a subvariety
$R_1$ of real dimension $2n-2$ with convex hull 
equal to the whole space $\mathcal{D}_n$. The Bloch sphere is an instance of this phenomenon
with $n=2$. So, even though the set of rank 1 density matrices is {\it small}, it can be used to generate the whole space of all density matrices of any rank. Since any point is obtained as convex sums of the vertexes (the pure states according to definition 4.2), this geometrical perspective allows to introduce the following abstract definition of {\it mixture} as a convex sum of abstract pure states.  

\begin{definition}[Abstract Mixture] A convex sum of abstract vectors or rank-1 matrices. 
\[
\rho^{mix} = \sum_{i} p_{i} \  \rho_{i}^{pure} =  \sum_{i} p_i \ \psi_i^\dag \psi_i.
\]
\end{definition}
\noindent However, like in the case of definition 4.2, these abstract definition of mixture has no operational content whatsoever. In fact, as we did with pure states, this abstract mixture must be clearly distinguished from operational mixtures which are a specific representation of an abstract mixture relative to a specific basis (see Appendix B). 
\begin{definition}[Operational Mixture] An abstract density matrix $\rho^{mix}$ can be specified in terms of an operational density matrix $[\rho^{mix}]_B$ relative to the basis $B$,
\[
[\rho^{mix}]_B =
\begin{pmatrix}
a&b\\c&d
\end{pmatrix}
\]
where $a, d \in \mathbb{R}$ and $b,c\in\mathbb{C}$, $\overline{b}=c$.
\end{definition}

To sum up, since in general abstract mixtures of rank $\neq 1$ cannot be represented as a unit vector, $|\psi \rangle$, there will exist no basis in which a mixed state will predict with certainty (probability = 1) a {\it yes-no} answer for a specific observable. And thus, while pure states are understood as providing maximal knowledge (definition 4.1), mixed states are seen as computing ignorance with respect to abstract vectors (definition 4.2). The problem is that abstract mixtures, conceived in terms of convex sums of abstract vectors, are not specified in terms of reference frames (or bases) and consequently have no operational content. It is only with the addition of a reference frame that it becomes possible for operational mixtures to provide a link with experience. As a consequence, the essential equivocity present in the notion of pure state is extended also to mixtures. Since Dirac and von Neumann the orthodox literature has uncritically confused these two distinct levels, which we have called `abstract' and `operational'. Even though these levels are irrelevant in the case of abstract mathematics they are fundamental when discussing within physics (see Appendix C).  

\smallskip 

It is interesting to notice that, apart from the original criticisms to the orthodox interpretation of mixtures discussed by Schr\"odinger already in 1936 \cite{Schr36}, there have been several expositions in the philosophical literature which continued, in different ways, addressing these problems. In a proceedings paper titled ``A Dilemma for the Traditional Interpretation of Quantum Mixtures'' published in 1972 Nancy Cartwright pointed out that  ``a dilemma which arises when one attempts to specify which decomposition is appropriate to a given physical situation.'' As she argued: 
\begin{quotation} 
\noindent {\small ``It is easy to show that any collection built up to accord with a particular decomposition of $W$ will yield the statistical results predicted by $W$. But the converse is not true. $W$ does not uniquely determine the constitution of the collection, since every mixed operator can be decomposed in more than one way. In the traditional view, however, no individual system can be in more than one pure state, so that the actual constitution of a physical collection can be described by one and only one decomposition.'' \cite[p. 251]{Cartwright72}}
\end{quotation} 
Cartwright concluded in her work that:
\begin{quotation} 
\noindent {\small ``not all mixed collections are composed of pure states; therefore it is not surprising that the ignorance interpretation faces difficulties in specifying what the pure states are on every occasion.'' \cite[p. 252]{Cartwright72}}
\end{quotation} 

Four years later, Bernard D'Espagnat would expose another internal inconsistency of the ignorance interpretation of mixtures. Following Landau and the atomist presupposition that the reduced matrices of a pure state can be interpreted as mixed states representing subsystems of the original pure state, D'Espagnat showed that one can derive a contradiction. Such reduced matrices simply cannot be interpreted in the orthodox manner as statistical mixtures of different pure states (see for a detailed derivation of the this inconsistency \cite[Chap. 6]{DEspagnat76}). In order to distinguish these new ``mixtures'', which cannot be interpreted in epistemic terms as convex sums of abstract vectors, D'Espagnat introduced the term ``improper''. These {\it improper mixtures} exposed a limit to the ignorance interpretation from within the pure-mixture account of matrices. By simply closing the circle and putting to play the metaphysical atomist understanding of pure states together with the ignorance interpretation of mixtures. Improper mixtures break the purity-mixture distinction from within SQM itself. Since there is no room for {\it improper mixtures} within SQM, while philosophers have learned to forget about them (there are less and less papers making reference to this uncomfortable improper-proper distinction), physicists have remained at safe distance from what they consider to be a purely philosophical debate and continue to uncritically apply the Landau-von Neumann interpretation of mixtures.\footnote{Notice that this is done even in the case of decoherence where it has been already demonstrated that there are deep problems with the application and interpretation of the notion of mixture \cite{Landsman95, Romano21, VassalloRomano23}.}

Finally, in \cite[p. 228]{Park88}, James Park shows that the ``ignorance interpretation of mixtures is flawed'' by extending some of the consequences of Schr\"odinger's probability relations \cite{Schr36} to quantum thermodynamics. Park considers a system $\mathcal{H}_1$ and its observer $X_1$, where $X_1$ has gathered sufficient data to compute the means for a quorum of observables of $\mathcal{H}_1$ and from them has concluded that $\rho_1$ has a specific form. According to the author, there are at least three conceptually distinct methods that could have been invoked to generate the ensemble that was presented to $X_1$: statical mixing, thermal equilibration and subensemble selection through correlations. $X_1$ cannot distinguish among the three methods even though he has the empirical means to measure every observable of $\mathcal{H}_1$.  Of particular interest to our works is section 2 in which the author analyzes the ambiguity of mixtures. Given it is completely natural to consider non-canonical ensembles to construct $\rho_1$:
\begin{quotation} 
\noindent {\small ``For $X_1$ this is nothing more than two mathematical expressions for the same quantum state $\rho_1$, which in itself completely summarizes all
statistics for all data sets $X_1$ can measure. In other words, for $X_1$
the distinct structures of the expansions themselves have no physical meaning
and are in fact undiscoverable. 

Against this background, the ignorance interpretation of the density operator as used in information-theoretic statistical thermodynamics seems shaky. [...] In the mathematics of quantum theory, multiple expansions like $\rho^{1} = \sum_{j} p_{j} |\psi_j\bra  \ket\psi_j |= \sum_{i}w_{i} |\phi_i\bra  \ket\phi_i |$ are quite normal, and therefore $\rho^1$ itself simply does not capture the idea that subjective probabilities have been assigned to an underlying list of pure states.'' \cite[p. 228]{Park88}}
\end{quotation} 

To conclude, it is also interesting to point out that while the operational approach to QM flourished ---together with quantum logic--- during the 1970s and 1980s, the use and application of mixtures became truly relevant during the 1990s when the operational definition of purity became to be regarded as ``too limited'' \cite[p. 419]{Hall13}. Since the notion of entanglement was orthodoxly defined in terms of the separability of systems into subsystems, the work of Landau became relevant. Unlike for vectors, the geometrical approach is able to provide a definition of separability and entanglement in terms of abstract density matrices (see \cite{BZ17}). In this context, a \emph{product state} is an abstract density matrix given by the tensor product of two abstract density matrices,  $\rho=\rho_1\otimes\rho_2$, where $\rho_1$ is a state in $\mathcal{H}_1$ and $\rho_2$ is a state in $\mathcal{H}_2$. The set of  \emph{separable states} in $\mathcal{H}_1\otimes \mathcal{H}_2$ is, by definition,  the closure (in trace norm) of  convex combinations of product states, $t_1\rho_{11}\otimes\rho_{21}+\dots+t_n\rho_{1n}\otimes\rho_{2n}$ with $t_1+\dots+t_n=1$ and $t_1,\dots,t_n\in\mathbb{R}_{\ge 0}$. Any state outside the set of \emph{separable states} is called an \emph{entangled state}. It is of course not surprising to find an essential difficulty in the contemporary debate about quantum entanglement which grounds its definition in the orthodox Dirac-von Neumann framework we have criticized in this work. Not surprisingly, these difficulties are also linked to the role of reference frames when considering the entanglement of states (see for a detailed discussion \cite{deRondeFMMassri24b, deRondeMassri22a}).

\section{Final Remarks}

Standard QM is grounded not only in an inconsistent definition of purity but also in an inconsistent extension of the notion of mixture derived from the (basis-independent) geometrical definition of pure state (i.e., as a geometrical abstract mixture) but interpreted in terms the (basis-dependent) operational definition of purity (i.e., as ignorance wrt certainty). As we have stressed these definitions can be found within the orthodox quantum physical literature and are completely independent of the interpretational debate that populates philosophical journals. Not only the notion of pure state is implied by contradictory definitions confused within the literature (section 5), also the notion of mixture remains both formally and conceptually inconsistent (section 6). While the operational definition of purity relates to the certain prediction of outcomes through the reference to a (preferred) {\it maximal basis}, the geometrical definition of purity rests on a purely abstract understanding of vectors devoid of any reference to bases and ---consequently--- detached from empirical content. As we have seen, the extension of pure states to density operators through the convex geometrical approach provides an abstract basis-independent definition of mixture which, in turn, is also related to the dual (inconsistent) definition of purity. Mixtures are understood as convex sums of trace-invariant abstract pure states, and also, as involving ignorance about the certainty implied by operational pure states. If we are willing to restore a rational analysis and debate about QM, the widespread distinction between pure and mixed states in the foundational and philosophical fields needs to be immediately abandoned and understood for what it is, a deep mistake of the past which has led both quantum physics and philosophy of QM into futile debates about pseudo-problems. 

\begin{center}
\section*{Appendix A: Operational states in Heisenberg's matrix formulation.}
\end{center}

In Heisenberg's formulation where the original valuation of matrices is linked to (non-binary) intensive values there is no distinction between pure and mixed states. All matrices in each and every basis provide a meaningful operational content related to a state of affairs. To fix ideas, let us provide an example in dimension 4. In this case, we can understand each and every point within the 4-simplex as referring to a particular laboratory. All these points, representing laboratories, are essentially equivalent from a physical perspective. 
\smallskip   
\begin{center}
\includegraphics[width=11em]{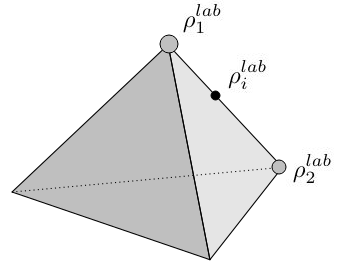}
\end{center}
If we now consider the operational level, it is clear that all bases provide meaningful content. While each abstract mixture can be naturally interpreted in terms of a laboratory, the choice of the basis specifies a sub-set of observables and their intensive values which in turn can be understood as a possible configuration of the measurement apparatuses within the lab, i.e. as particular experimental contexts. For each lab we can think of different experimental arrangements in each of which there is a specific set of observables with definite intensive values. Such intensive values are of course given through Born's rule reinterpreted now ---beyond particles and measurement outcomes--- as quantifying intensities. Thus, every abstract matrix, $\rho^{lab}_i$, representing a laboratory gives rise to many different experimental contexts within that lab, $[\rho^{lab}_i]_{B_j}$, each one of them specified by a particular basis $B_j$.
\begin{center}
\includegraphics[width=17em]{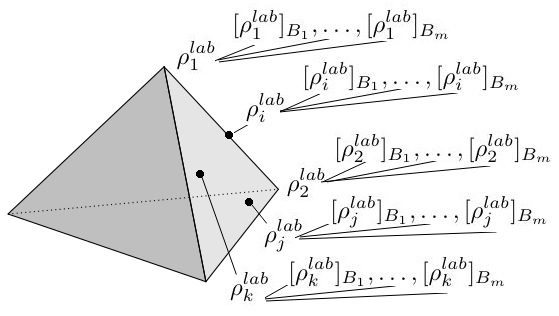}
\end{center}
In Heisenberg's formulation all matrices and all bases stand on equal footing, all laboratories and experimental contexts are ---as required in physics--- essentially equivalent. There are no preferred labs, and there are no preferred bases.

\begin{center}
\section*{Appendix B: The `abstract-operational' and `purity-mixture'\\distinctions in the Dirac-von Neumann formulation.}
\end{center}

Let us begin by recalling the difference between what we have labeled `the abstract' and `the operational' levels of representation of states which are implicitly addressed within the standard Hilbert space formulation of QM. Let ${\cal H}$ be a $n$-dimensional abstract Hilbert space. The choice of an ordered basis $B$ for ${\cal H}$ defines an isomorphism between ${\cal H}$ and $\mathbb{C}^n$. In particular, each abstract vector $v\in {\cal H}$ can be written as  an array of numbers $[v]_B\in \mathbb{C}^n$. It is important to notice that a change of ordered basis $B'$ in  ${\cal H}$, changes the representation of $[v]_B\in\mathbb{C}^n$ to another vector $[v]_{B'}$ in $\mathbb{C}^n$. Thus, {\it the same} vector $v$ possesses a multiplicity of {\it different} bases dependent representations, $[v]_{B}, [v]_{B'}, [v]_{B''}, ..., [v]_{B^n}$. In particular, given a vector $v\in {\cal H}$, it is always possible to choose an ordered basis such that the corresponding vector in $\mathbb{C}^n$ is equal to $(1,0,\dots,0)$. As we have noticed this is a kernel feature within the Dirac-von Neumann formulation in order to consider pure states in which binary certainty is accomplished. It is essential to recognize that while ---since there is a complete mathematical equivalence between these levels--- the distinction is completely unimportant for mathematicians, the distinction is kernel for physicists when attempting to represent any {\it state}. For states in physics necessarily presuppose the specification of a {\it reference frame} or {\it basis}. In fact, the {\it valuation} of the state of a system ---namely, the values of the properties of the system in a specific situation--- can be only given as relative to a {\it reference frame}. The position or velocity of a system only makes sense when considered as relative to a reference frame. A state with no mention to a specific reference frame in physics is simply meaningless. However, as we have seen in section 5, in the Dirac-von Neumann orthodox formulation pure states and mixtures are implicitly considered in both abstract and operational levels. While the abstract purity-mixture distinction is linked to the rank of matrices, the operational purity-mixture distinction is related to the predictive certainty or ignorance about measurement outcomes (or states). The extension of abstract pure states to abstract mixtures is grounded on the fact that, from a mathematical perspective, rank 1 matrices can generate the whole space of matrices of any rank. Continuing with our previous example in dimension 4, ssume that we choose the black point in the 4-simplex, $\rho^{mix}_i$, that lies in the line in between $\rho^{pure}_1$ and $\rho^{pure}_2$ in the following figure,
\smallskip
\begin{center}
\includegraphics[width=11em]{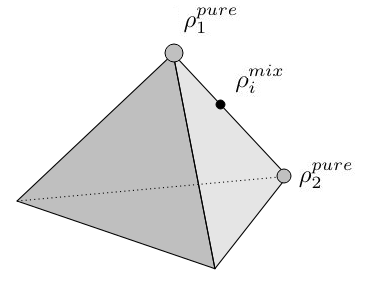}
\end{center}
Then, $\rho^{mix}_i$ is a rank $2$ diagonal matrix in $\mathbb{C}^{4\times 4 }$ that represents a rank $2$ abstract matrix. It is obvious that any abstract matrix which is in the line between the abstract pure states $\rho^{pure}_1$ and $\rho^{pure}_2$ will be obtainable as a convex sum of them:
\[
\rho^{mix}_i= t \ \rho^{pure}_1 + (1-t) \ \rho^{pure}_2
\]
The operational purity level requires then the specification of a basis which in the Dirac-von Neumann formulation is given by the preferred {\it maximal basis} which no other than the diagonal basis. In our example these are the operational matrices $[\rho^{pure}_1]_{B_{diag}}$ and $[\rho^{pure}_2]_{B_{diag}}$, which in explicit terms read:
\[
[\rho^{pure}_1]_{B_{diag}} =
\begin{pmatrix}
0&0&0&0\\
0&1&0&0\\
0&0&0&0\\
0&0&0&0\\
\end{pmatrix} 
\ \ \ \ \ \
[\rho^{pure}_2]_{B_{diag}} =
\begin{pmatrix}
0&0&0&0\\
0&0&0&0\\
0&0&1&0\\
0&0&0&0\\
\end{pmatrix}
\]
Thus, we can write: 
\[
\rho^{mix}=
\begin{pmatrix}
0&0&0&0\\
0&t&0&0\\
0&0&1-t&0\\
0&0&0&0\\
\end{pmatrix}=
t
\begin{pmatrix}
0&0&0&0\\
0&1&0&0\\
0&0&0&0\\
0&0&0&0\\
\end{pmatrix}+
(1-t)
\begin{pmatrix}
0&0&0&0\\
0&0&0&0\\
0&0&1&0\\
0&0&0&0\\
\end{pmatrix}.
\]

To sum up, in the $4$-dimensional simplex there are only 4 points which represent abstract pure states, namely, $\rho^{pure}_1, \rho^{pure}_2, \rho^{pure}_3$ and $\rho^{pure}_4$, and for each of these points there is only one preferred basis ---namely, the diagonal basis--- which has operational content. Thus,  in the $4$-dimensional simplex we obtain only 4 operational states $[\rho^{pure}_1]_{B_{diag}}, [\rho^{pure}_2]_{B_{diag}}, [\rho^{pure}_3]_{B_{diag}}$ and $[\rho^{pure}_4]_{B_{diag}}$ which are ``truly quantum''. 
It becomes clear that ``purity'' means nothing else than binary certainty. It is important to remark that even though the abstract points are not the same as the operational points, their number is exactly the same for the restriction is filtered by a single preferred basis. Furthermore, mixed states, which are the remaining points in $n$-simplex, must be understood as combinations of only these 4 points (i.e.,the  pure states). As always, quantum superpositions remain essentially problematic within the orthodox literature (see \cite{deRonde18}). 
\begin{center}
\includegraphics[width=25em]{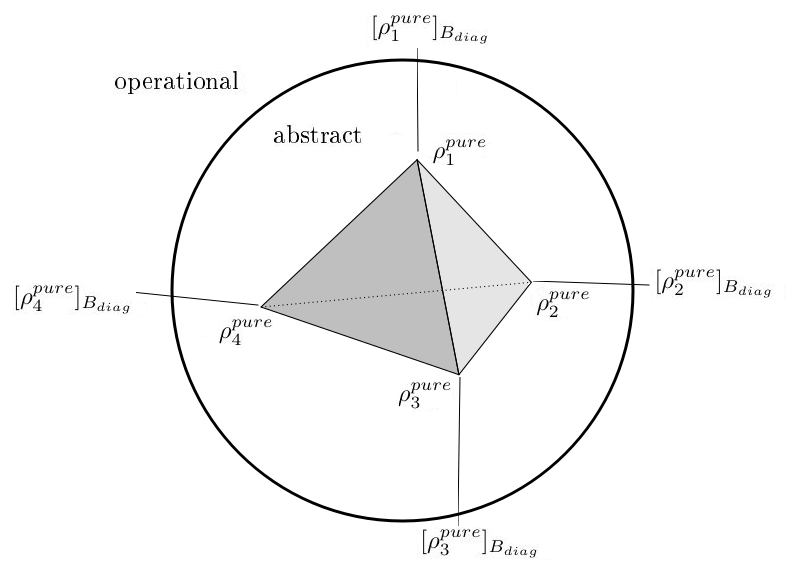}
\end{center}
 
\begin{center}
\section*{Appendix C: Quantifying operational states in the Heisenberg and the Dirac-von Neumann formulations.}
\end{center}

If we now compare, in the $n$-simplex, the operational states in the Dirac-von Neumann and in the Heisenberg formulations we obtain that while in the first case there are only $n$ truly quantum operational states, in the matrix formulation we have an infinite number of states. In this latter case, every point in the $n$-simplex represents a lab, $\rho^{lab}_i$, and for each one of these points every possible basis represents a specific fully operational experimental context $[\rho^{lab}_i]_{B_j}$. All these labs and basis dependent contexts are of course completely equivalent from an operational viewpoint. 
\smallskip 

\begin{center}
\includegraphics[width=25em]{d-vn.png}
\includegraphics[width=22em]{h.png}
\end{center}
Even though the difference is quite obvious by comparing the figures of the 4-simplex, we can quantify the operational states that appear in the different formulations  through dimensional analysis. In the vectorial formulation of Dirac and von Neumann the space of pure states consists of $n$ finite elements giving a geometrical structure of dimension $0$. The specification of a basis does not change the dimension of the set of pure states, which continues to be $0$.  On the contrary, in Heisenberg's matrix formulation we need to consider the space of density matrices  $\mathcal{D}$ which has dimension $n^2-1$ and the space of orthonormal frames $\mathcal{F}^\bot$ of dimension $n^2$,
\begin{align*}
\mathcal{D}&:=\mathcal{D}({\cal H}) = \{\rho\in B(\mathcal{H})\,\colon\, \rho\text{ is a density matrix}\} \\
\mathcal{F}^\bot: &= \mathcal{F}^\bot({\cal H}) = \left\{ 
(v_1,\dots,v_n)\in {\cal H}^n\,\colon\, \{v_1,\dots,v_n \} \text{ is a orthonormal basis}
\right\}
\end{align*}

\noindent{\bf Definition. }
An element $(\rho^{lab},B)\in \mathcal{D}\times \mathcal{F}^\bot$ denoted $[\rho^{lab}]_B$
is called {\it operational density matrix} and an element $\rho^{lab}\in\mathcal{D}$ is called an
{\it abstract density matrix}.

\

\noindent Now that we have all the necessary elements, we can count the dimension of the space of operational density matrices in Heisenberg's formulation, which is $\mathcal{D}\times \mathcal{F}^\bot$.
The space of abstract density matrices $\rho^{lab}$ has dimension $n^2-1$ and for each one of these matrices we can choose a different basis $B\in\mathcal{F}^\bot$. 
Hence, $\mathcal{D}\times \mathcal{F}^\bot$ has dimension $2n^2-1$ which represents the
different ways of choosing a density matrix in a specific basis $[\rho^{lab}]_B$. 
Summarizing, while the space of operational density matrices in the Dirac-von Neumann formulation has dimension $0$, in Heisenberg's formulation it has dimension $2n^2-1$.




\section*{Acknowledgements} 

This work was partially supported by the following grants: Project PIO-CONICET-UNAJ (15520150100008CO) ``Quantum Superpositions in Quantum Information Processing''. UFSC Processo 202013133. The authors state that there is no conflict of interest. We want to thank two anonymous reviewers for their comments and criticisms.

\end{document}